\documentclass[titlepage,twoside,12pt]{article}
\usepackage{amssymb}
\usepackage{amsfonts}
\begin{document}

{\bf \Large Velocity addition in Special Relativity and in Newtonian Mechanics are
isomorphic} \\ \\

Elem\'{e}r E Rosinger \\
Department of Mathematics \\
and Applied Mathematics \\
University of Pretoria \\
Pretoria \\
0002 South Africa \\
eerosinger@hotmail.com \\ \\

{\bf Abstract} \\

In the one dimensional case, velocity addition in Special Relativity and in Newtonian
Mechanics, respectively, are each a commutative group operation, and the two groups are {\it
isomorphic}. There are {\it infinitely} many such isomorphisms, each indexed by one positive
real parameter. \\ \\

{\bf 1. Velocity addition in Special Relativity} \\

Let $c > 0$ be the velocity of light in vacuum. Then, as is well known, Angel, in the case of
uniform motion along a straight line, the special relativistic addition of velocities is given
by \\

(SR) $~~~~~~ u * v ~=~ ( u + v ) / ( 1 + u v / c^2 ),~~ u, v \in ( -c, c)$ \\

thus the binary operation $*$~ acts according to \\

$~~~~~~ * : ( -c, c ) \times ( -c, c ) \longrightarrow ( -c, c )$ \\

It follows immediately that \\

1) $*$~ is associative and commutative \\

2) $u * v * w ~=~ ( u + v + w + u v w / c^2 ) / ( 1 + ( u v + u w + v w ) / c^2 )$ \\

for $u, v, w \in ( -c, c)$ \\

3) $u * 0 ~=~ 0 * u ~=~ u,~~ u \in ( -c, c)$ \\

4) $u * ( -u ) ~=~ ( -u ) * u ~=~ 0,~~ u \in ( -c, c)$ \\

5) $\partial / \partial u ( u * v ) ~=~ ( 1 - v^2 / c^2 ) / ( 1 + u v / c^2 )^2 > 0,~~
                                                                   u, v \in ( -c, c)$ \\

6) $\lim_{u, v \to c}~ u * v ~=~ c,~~~~ \lim_{u, v \to -c}~ u * v ~=~ -c$ \\

Therefore \\

7) $(~ ( -c, c ), * ~)$ is a commutative group with the neutral element $0$, while $-u$ is the
inverse element of $u \in ( -c, c)$ \\ \\

{\bf 2. Velocity addition in Newtonian Mechanics} \\

As is well known, in the case of uniform motion along a straight line, the addition of
velocities in Newtonian Mechanics is given by \\

(NM) $~~~~~~ x + y,~~ x, y \in \mathbb{R} $ \\

thus it is described by the usual additive group $( \mathbb{R}, + )$ of the real numbers, a
group which is of course commutative, with the neutral element $0$, while $-x$ is the
inverse element of $x \in \mathbb{R}$. \\ \\

{\bf 3. Isomorphisms of the two groups} \\

8) $(~ ( -c, c ), * ~)$ and $( \mathbb{R}, + )$ are isomorphic groups through the mappings \\

8.1) $~~~ \alpha : ( -c, c) \longrightarrow \mathbb{R}$, where \\

$~~~~~~ \alpha ( u ) ~=~ k \ln ( ( c + u ) / ( c - u ) ),~~ u \in ( -c, c)$ \\

and \\

8.2) $~~~ \beta : \mathbb{R} \longrightarrow ( -c, c)$, where \\

$\beta ( x ) ~=~ c ( e^{ x / k } - 1 ) / ( e^{ x / k } + 1 ),~~ x \in \mathbb{R}$ \\

with \\

8.3) $~~~ k ~=~ c^2 \alpha^\prime ( 0 ) > 0$ \\ \\

{\bf Proof of 8)} \\

Let us first find $\alpha$. According to the standard definition of group homomorphism, we
have  \\

$~~~~~~ \alpha ~\mbox{group homomorphism} ~\Leftrightarrow~ \alpha ( u * v ) = \alpha ( u ) +
  \alpha ( v ),~~ u, v \in ( -c, c)$ \\

Thus it follows that \\

$~~~~~~ \alpha ( u * v ) - \alpha ( u ) ~=~ \alpha ( v ),~~ u, v \in ( -c, c)$ \\

and since the right hand term does not depend on $u$, we conclude that neither does the left
hand term. Consequently, assuming that $\alpha$ has a derivative on its domain of definition
$( -c, c)$, we obtain \\

$~~~~~~ d/du ~( \alpha ( u * v ) - \alpha ( u ) ) ~=~ 0,~~ u, v \in ( -c, c)$ \\

or in view of (SR) and 5), the relation follows \\

$~~~~~~ \alpha^\prime ( ( u + v ) / ( 1 + u v / c^2 ) ) ( ( 1 - v^2 / c^2 ) /
                     ( 1 + u v / c^2 ) ) ~=~ \alpha^\prime ( u ) $ \\

for $u, v \in ( -c, c)$ \\

Taking now $u = 0$, one obtains \\

$\alpha^\prime ( v ) ( 1 - v^2 / c^2 ) ~=~ \alpha^\prime ( 0 ),~~ v \in ( -c, c)$ \\

or \\

$\alpha^\prime ( v ) ~=~ c^2 \alpha^\prime ( 0 ) / ( c^2 - v^2 ),~~ v \in ( -c, c)$ \\

Thus, since $\alpha ( 0 ) ~=~ 0$ results form the fact that $\alpha$ is assumed to be a group
homomorphism, one obtains \\

$ \begin{array}{l}
     \alpha ( u ) ~=~ \alpha ( 0 ) + c^2 \alpha^\prime ( 0 ) \int_0^u dv / ( c^2 - v^2 ) ~=~
     c^2 \alpha^\prime ( 0 ) \int_0^u dv / ( c^2 - v^2 ) ~=~ \\ \\
     =~ c^2 \alpha^\prime ( 0 ) \ln ( ( c + u ) / ( c - u ) ),~~ u \in ( -c, c)
  \end{array} $ \\

in other words, 8.1) and 8.3). And since obviously the resulting $\alpha$ in 8.1) is a
{\it bijective} mapping, it follows that it is not only a group homomorphism, but also a group
isomorphism. In this way, its inverse mapping $\beta = \alpha^{-1}$ exists and it is also a
group isomorphism. Finally, a simple computation based on 8.1) will then give 8.2). \\ \\

{\bf 4. Note} \\

The special relativistic addition $*$~ of velocities in (SR) is in fact well defined not only
for pairs of velocities \\

$~~~~~~ ( u, v ) \in ( -c, c) \times ( -c, c)$ \\

but also for the {\it larger} set of pairs of velocities \\

$~~~~~~ ( u, v ) \in [ -c, c ] \times [ -c, c ],~~ u v \neq - c^2 $ \\

This corresponds to the fact that in Special Relativity the velocity $c$ of light in vacuum is
supposed to be attainable. \\

On the other hand, the Newtonian addition $+$~ of velocities (NM) does of course only make
sense physically for \\

$~~~~~~ ( x, y ) \in \mathbb{R} \times \mathbb{R}$ \\

since infinite velocities are not supposed to be attainable physically. \\

As for the group isomorphisms $\alpha$ and $\beta$, they only generate mappings between pairs
of velocities in \\

$~~~~~~ ( -c, c) \times ( -c, c) ~\stackrel{\alpha} \longrightarrow~
                                             \mathbb{R} \times \mathbb{R}$ \\

and \\

$~~~~~~ \mathbb{R} \times \mathbb{R}~\stackrel{\beta} \longrightarrow~
                                                       ( -c, c) \times ( -c, c)$ \\

thus they do {\it not} cover the cases of addition $u * v$ of special relativistic velocities
$u = -c$ and $v < c$, or $-c < u$, and $v = c$. \\

Consequently, in spite of the group isomorphisms $\alpha$ and $\beta$, there is an {\it
essential} difference between the addition of velocities in Special Relativity, and on the
other hand, Newtonian Mechanics. Indeed, in the latter case, the addition $+$~ is defined on
the {\it open} set $\mathbb{R} \times \mathbb{R}$, while in the former case the addition $*$~
is defined on the set \\

$~~~~~~ \{~ ( u, v ) ~|~ -c \leq u, v \leq c,~~ u v \neq -c^2 ~\}$ \\

which is {\it neither open, nor closed}. \\ \\

{\bf 5. The uniqueness of the velocity addition in \\
\hspace*{0.45cm} Special Relativity} \\

In Benz, it has recently be shown that under very general and mild conditions, the formula
(SR) is {\it uniquely} determined, even if one starts with motions not along a straight line,
but in arbitrary, thus possibly infinite dimensional pre-Hilbert spaces as well. \\ \\

{\bf Reference} \\

1. Angel, Roger B : Relativity, The Theory and its Philosophy. Pergamon, New York, 1980 \\

2. Benz, W : A characterization of relativistic addition. Abh. Math. Sem. Hamburg, vol. 70,
2000, pp. 251-258

\end{document}